\documentclass[showpacs,pre,floatfix,twocolumn]{revtex4-1}
\usepackage{amsmath,amsfonts,amssymb}
\usepackage{graphicx} 
\usepackage{color} 
\begin{document}

\title{Effect of dynamic and static friction on an asymmetric granular piston}


\author{Julian Talbot and Pascal Viot}

\affiliation{Laboratoire de Physique Th\'eorique de la Mati\`ere Condens\'ee, UPMC, CNRS  UMR 7600,
 4, place Jussieu, 75252 Paris Cedex 05, France
}

\email{talbot@lptmc.jussieu.fr}
\email{viot@lptmc.jussieu.fr}
\date{\today}
\begin{abstract}
We investigate the influence of dry friction on an asymmetric,  granular
piston of mass $M$ composed of two materials  undergoing inelastic collisions with
 bath particles of mass $m$. Numerical simulations of the Boltzmann-Lorentz equation
reveal the existence of two scaling regimes depending on the strength of
friction. In the large friction
limit, we introduce  an exact model giving the  asymptotic
behavior of the Boltzmann-Lorentz equation. For small
friction and for large mass ratio $M/m$, we derive a Fokker-Planck
equation for which the exact solution is also obtained. Static friction attenuates the motor effect and
results in a discontinuous velocity distribution.
\end{abstract}

\maketitle
\section{Introduction}\label{intro}
An adiabatic piston separating two compartments of gases is a widely studied model in statistical physics. The initial interest arose from the observation 
that the equilibrium state cannot be predicted by application of the First and Second laws of thermodynamics \cite{callen60,Piasecki1999,Gruber1999,Gruber2002,Gruber2003}. It was shown that dynamics
contains different  time scales before  the system reaches equilibrium (for finite-size compartments) or a steady state where
the piston acquires a non-zero drift velocity (for infinite compartments) \cite{Gruber2006}.

The granular version of the system, where the gas particles undergo dissipative collisions, also displays interesting behavior. 
Brito et. al \cite{Brito2005} showed that the piston eventually collapses to one side and Brey and Khalil \cite{PhysRevE.82.051301} showed that the steady state 
is characterized by equal cooling rates in the two compartments. 

The model was investigated in the context of a granular motor by Costantini et al. \cite{costantini:061124}. They considered a
piston composed of two different materials and showed that
fluctuations on the right and left sides result in noise rectification that can be converted into mechanical work. 
When the bath density is low, the appropriate kinetic description of the system is  the Boltzmann-Lorentz equation. 
However, an exact solution cannot be obtained even for this simple model. Costantini {\it et al.} \cite{Costantini2008} proposed an ansatz
of the velocity distribution, where parameters are obtained by calculating successive moments of the kinetic equation. Comparisons with
numerical simulations showed that the approach is reasonable, but it fails in the limit of large piston mass (the Brownian limit).
Talbot {\it et al.} \cite{PhysRevE.82.011135} introduced  a mechanical treatment that gives an exact  expression for
the drift velocity in the Brownian limit.


Recently, Eshuis et al. \cite{PhysRevLett.104.248001} presented the first experimental realization of a macroscopic, rotational ratchet in a granular gas. 
The device,  which consists of four vanes, is reminiscent of that imagined by 
Smoluchowski \cite{smolu12,feymann63}. When a soft coating was
applied to one side of each vane, a motor effect was observed above a critical granular temperature.
While this was the first experimental realization of a granular motor, similar Brownian ratchets exist in many diverse applications, e.g., photovoltaic devices and biological motors; 
See \cite{Reimann2002,RevModPhys.69.1269,vandenBroekM.2009}.
All of these motors share the common features of non-equilibrium conditions and spatial symmetry breaking.
Several recent theoretical studies of idealized models of granular motors, which use a Boltzmann-Lorentz description
\cite{Cleuren2007,Cleuren2008,Costantini2008,Costantini2009,PhysRevE.82.011135}, confirm that the motor effect is particularly pronounced
when  the device is constructed from two different materials, as was the case in the recent experiment \cite{PhysRevLett.104.248001}.
The existing theories, however, predict a motor effect for any temperature of the granular gas while in the
experiment the phenomenon is only observed if the bath temperature is sufficiently large.

Friction likely plays an important role in the experiment \cite{PhysRevLett.104.248001}  as it does in other systems with stochastic dynamics. The first theoretical studies date from
2005 when de Gennes \cite{Gennes2005} and Hayakawa \cite{Hayakawa2005} addressed the effect of dry (Coulombic) friction 
on Brownian motion. Subsequently, Kawarada and Hayakawa \cite{JPSJ.73.2037} showed that the signature of Coulombic friction is an exponentially 
decaying velocity distribution function. 
Menzel and Goldenfeld \cite{PhysRevE.84.011122} studied a Fokker-Planck equation and noted a formal connection to the Schr\"odinger equation 
for the quantum mechanical oscillator with a delta potential.  Mauger\cite{Mauger2006} showed  that the Coulomb friction is responsible
for an exponential decay of the velocity distribution when dynamics is described by a Fokker-Planck equation.
Touchette and coworkers \cite{Touchette2010,Baule2010,Baule2011}  obtained a solution of a model with dry friction and viscous damping. 
Experimental studies have examined droplets on non-wettable surfaces subjected to an asymmetric lateral vibration \cite{A.Buguin2006},
 as well as the biased motion of a water drop on a tilted surface subject 
to vibration \cite{Chaudhury2008,Goohpattader2009,Mettu2010}. 

Recently, we used numerical simulation and kinetic theory to examine the effect of dynamic friction on a chiral rotor within the framework of the Boltzmann-Lorentz equation 
\cite{PhysRevLett.107.138001}. 
The numerical simulations revealed the existence 
of two scaling regimes at low and high bath temperatures. For large piston masses and small friction the model can be mapped onto a Fokker-Planck equation that can be solved analytically. We also obtained
analytic solutions for the mean  velocity and the velocity distribution function in the limit of large friction.   
The purpose of the present article is to present a complete analysis of the effect of dynamic and, for the first time, static friction on 
the asymmetric granular piston.

In Sec. \ref{sec:model}, we introduce the model of a granular piston with friction. We perform
Monte Carlo simulations in Sec.\ref{sec:numsim}. A time scale analysis presented in Sec. \ref{sec:tsanalysis} suggests that the model can be solved in two limiting cases.
In Sec.\ref{sec:ikm}, we first consider 
the high friction limit by  introducing the independent kick model and compare the exact solution of the model with numerical simulations of the BL equation. In the Brownian limit 
and in the small friction limit, we show that the Fokker-Planck equation, for which analytical solutions can be obtained, provides an accurate description of the BL equation. 
In Sec.\ref{sec:static}, we generalize our study by including the effect of static friction and we briefly conclude in Sec.\ref{sec:conclusion}.

\section{The model}\label{sec:model}

\begin{figure}[t]
\centering
\resizebox{9cm}{!}{\includegraphics{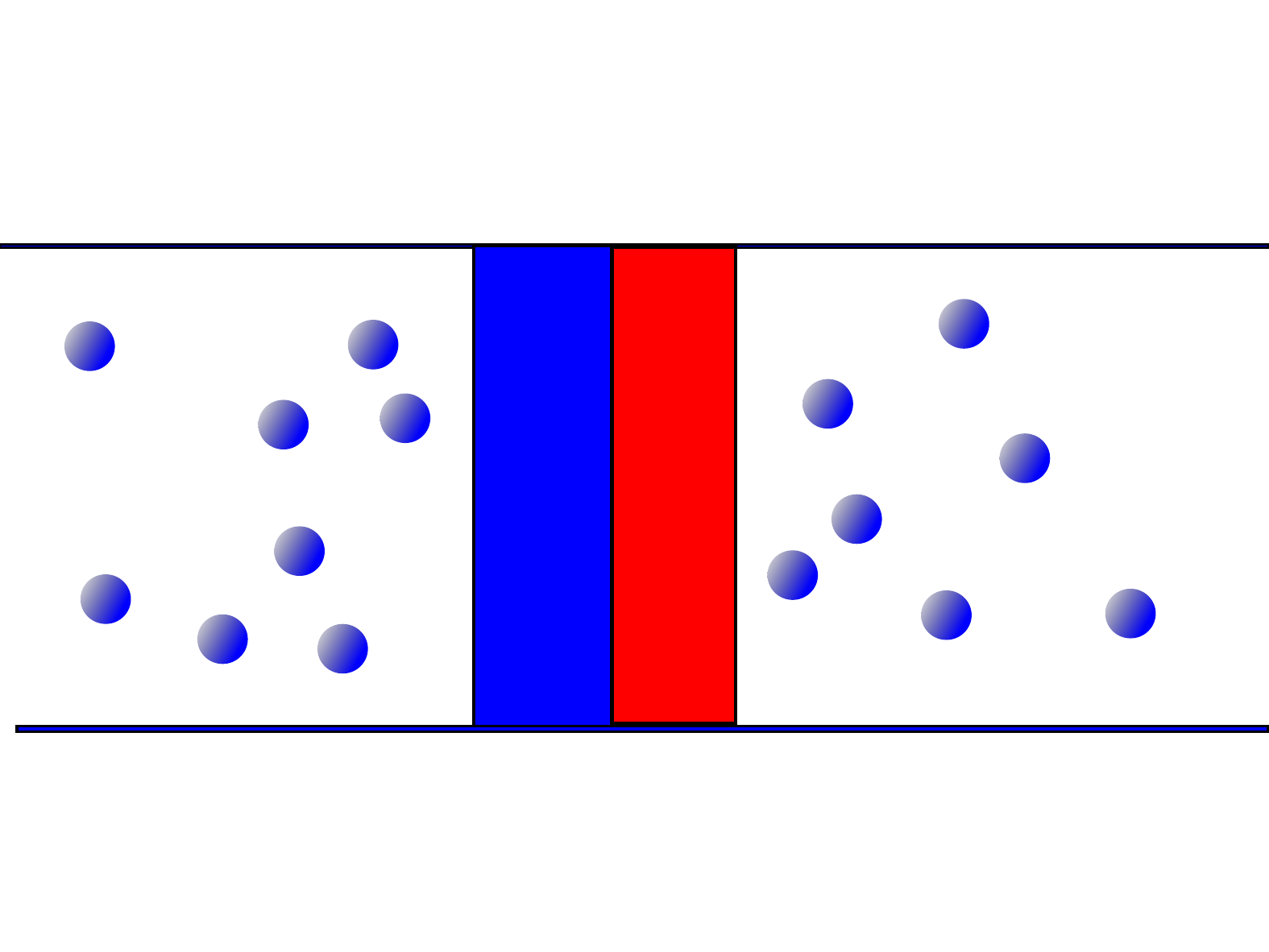}}
\caption{The asymmetric granular piston in a bath of thermalized particles.}
\label{fig:piston}
\end{figure}

An infinite cylinder filled with a monodisperse gas  
composed of particles of mass $m$ is separated into two
compartments by a granular piston of mass $M$ and of vertical length $L$ that is composed of two materials
characterized by the coefficients of restitution $\alpha_+$ and $\alpha_-$. The piston is constrained to move along the
symmetric axis of the container and undergoes collisions with the gas
particles. In addition, a frictional force of constant strength $F$, which acts to oppose the motion of
the piston, is present.  In the absence of the bath particles the equation of motion of the piston is 
\begin{equation}
M\frac{dV}{dt}=-\sigma(V)F
\end{equation}
where 
\begin{equation}
  \sigma(V) = \left\{ \begin{array}{rl}
 +1 &\mbox{ if $V>0$} \\
 0 &\mbox{ if $V=0$} \\
 -1 &\mbox{ if $V<0$} 
       \end{array} \right.
\end{equation}
Due to the translational invariance along perpendicular axis of the container and
assuming that the radius of the cylinder is sufficiently, boundary effects can be neglected. Consequently,
the collision rules only involve the velocity component
along the the cylinder axis. If $v$ and $V$ are the pre-collisional values, the post-collisional velocities of the piston and gas particles are
\begin{equation}\label{eq:colalpha}
 V'_{\alpha_{\pm}}= V + \frac{1+\alpha_{\pm}}{1+\mu}(v-V)
\end{equation}
\begin{equation}\label{eq:colalpha2}
 v'_{\alpha_{\pm}} = v - \frac{\mu (1+\alpha_{\pm} )}{1+\mu}(v-V)
\end{equation}
where  $\mu = \frac{M}{m}$ is the mass ratio and $\alpha_+$ ($\alpha_-$) is selected if the collision occurs on the left (right) hand side of the piston, i.e. if $v-V>0$ ($v-V<0$).

Pre-collisional (or restituting) velocities $V''_{\alpha_{\pm}}, v''_{\alpha_{\pm}}$ can be
obtained from Eqs.(\ref{eq:colalpha}-\ref{eq:colalpha2})  by replacing
$\alpha_{\pm}$ with $\alpha_{\pm}^{-1}$.

The kinetic properties of the piston are described  by means of the Boltzmann-Lorentz
equation. Let us denote  $f(V;t)$ the probability density  of
finding the piston moving with velocity $V$ and $\phi (v)$ represents the
velocity distribution of the bath particles
at time $t$, one has
\begin{equation}
\frac{\partial}{\partial t} f(V;t)-F\frac{\sigma(V)}{M}f(V;t) = J[\phi,f]
\end{equation}
where $J[\phi,f]$ is the collision operator expressed as 
\begin{align}&J[f,\phi]=
\rho L\int_{-\infty}^{\infty} dv \,
|v-V| [\theta(v-V) \frac{f(V''_{\alpha_{+}}
;t)}{\alpha_{+}^2}
\nonumber\\&\left.\phi (v'')
 + \theta(V-v) \frac{f(V''_{\alpha_{-}} ;t)}{\alpha_{-}^2}\phi (v'') \right]- \rho L\nu(V)f(V;t)
\end{align}
where $\theta(u) $ is the Heaviside function and
\begin{equation}
\rho L\nu(V)=\rho L \int_{-\infty}^{\infty} dv |v-V|\phi (v)
\end{equation}
is the collision rate of bath particles with (both sides of) the piston moving with a velocity $V$. 
Note that the Boltzmann-Lorentz equation neglects recollisions.  This assumption is valid if  the bath density is low and when
the mass of the piston is larger than the mass of bath particles. In addition,   the bath distribution, $\phi(v)$, is assumed stationary and symmetric
such that $\langle v\rangle=0$.

By using appropriate changes of variables\cite{piasecki:051307}, the
kinetic equation can be rewritten as
\begin{align}\label{eq:BLy}
&\frac{1}{\rho L}\frac{\partial}{\partial t}
f(V;t)-\frac{F\sigma(V)}{M\rho L} \frac{\partial}{\partial V} 
f(V;t) = \int_0^{\infty}dy\;y\nonumber\\
& \left[ f\left(V-\frac{1+\alpha_{+}}{1+\mu}y;t\right)\phi\left(V
+\frac{\mu-\alpha_{+}}{1+\mu}y\right)\right.\nonumber \\
&+\left.\;f\left(V+\frac{1+\alpha_{-}}{1+\mu
} y;t\right)\phi\left(V
-\frac{\mu-\alpha_{-}}{1+\mu}y\right)\right]\nonumber
\\
&-f(V;t)\int_0^{\infty}dy\;y(\phi(V+y)+\phi(V-y))
\end{align}

It is convenient to introduce the reduced variables $F^*=F/(\rho L T)$ and $V^*=\sqrt{m/T}V$. With this choice 
the  average drift velocity only depends  on $F^*$, $M/m$ and $\alpha_{\pm}$. 
In an experiment, the frictional force depends on the physical properties of
the motor and is not easily changed. On the other hand, the granular temperature
of the bath particles can be varied simply by increasing or
 decreasing the vibration amplitude or frequency of the mechanical shaker driving
the granular gas particles. 

Where possible we give analytic results for a general bath particle velocity distribution, $\phi(v)$. 
For illustrative purposes, as well as to test the theory by comparison with numerical simulation of the BL equation,  we will use a Gaussian distribution:
\begin{equation}
 \phi(v)=\sqrt{\frac{m}{2\pi T}}\exp\left(-\frac{mv^2}{2T}\right)
\end{equation}

\section{Numerical simulation}\label{sec:numsim}
\begin{figure}[t]
\resizebox{8.0cm}{!}{\includegraphics{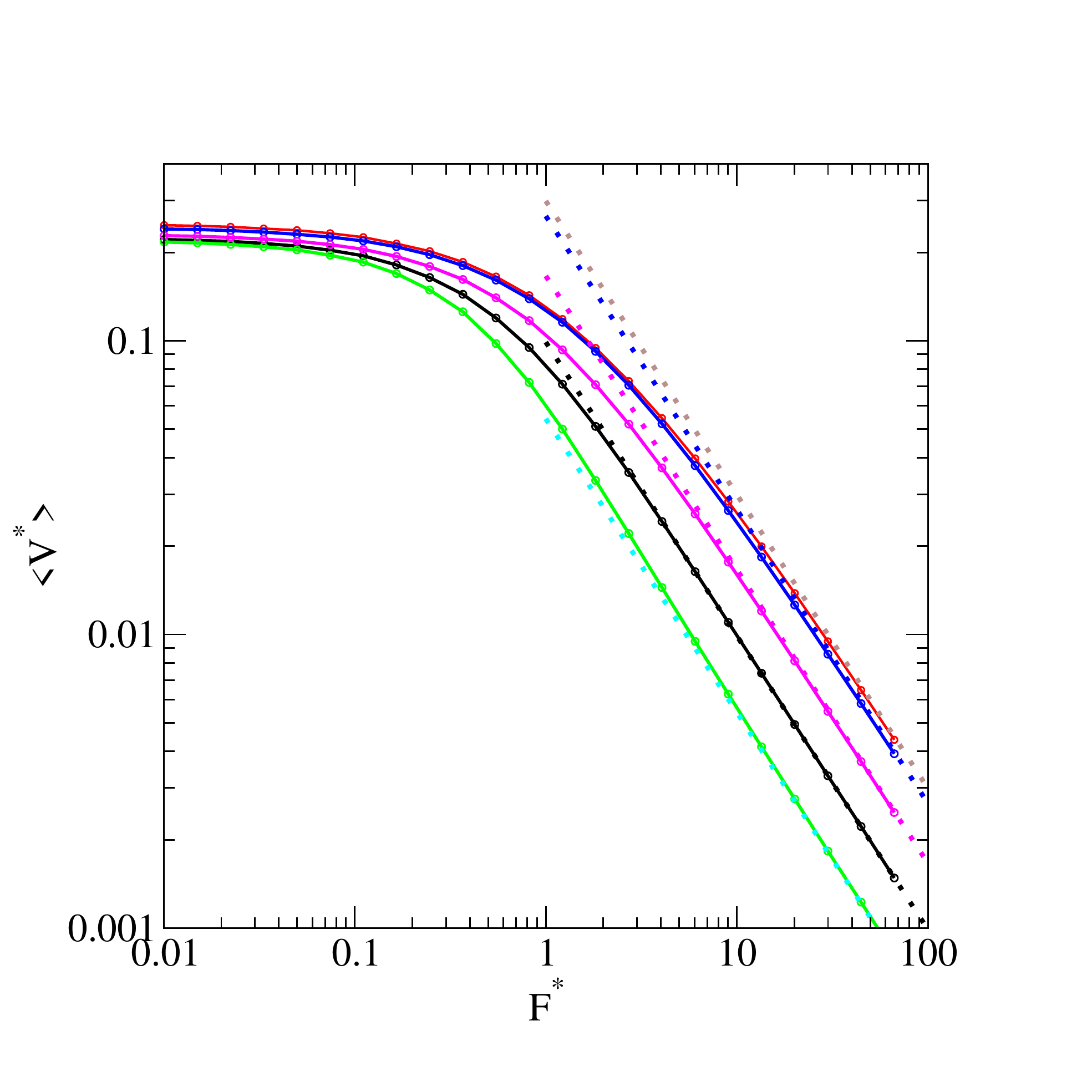}}
 \caption{(Color online) Log-Log plot of the dimensionless mean velocity $\langle
V^* \rangle$ of 
an asymmetric  granular piston versus the dimensionless friction force $F$: 
$\alpha_+=1$, $\alpha_-=0$, with
different mass ratio 
$M/m=1,2,5,10,20 $.  The dashed curves correspond to the analytical expression
of the
large friction model.}
 \label{fig:velocity_fric}
 \end{figure}

We performed numerical simulations of the Boltzmann-Lorentz equation
using the Gillepsie method \cite{TV06} for different mass ratios  and  for a large range
of dry friction. 
The algorithm generates collision events separated by exponentially distributed
waiting intervals. Specifically, the  probability that no event (collision) occurs in the time interval $(0,\Delta t)$ is given by
\begin{equation}
 P(\Delta t)=\exp\left(-\rho\int_0^{\Delta t}\nu(t')dt'\right)
\end{equation}
In the present application the mean collision flux $\rho\nu(t)$ is time dependent as the piston decelerates between collisions. 
A collision time, $\Delta t$, is generated by solving (numerically) the implicit equation
\begin{equation}
 \ln(\xi)=-\rho\int_0^{\Delta t}\nu(t')dt'
\end{equation}
where $0<\xi<1$ is a uniformly distributed random number. The system time is incremented by $\Delta t$ and the collision 
is performed by sampling a velocity of the colliding bath particle using the imposed velocity distribution $\phi(v)$ and
updating the piston's velocity using the collision rule Eq. (\ref{eq:colalpha}).  Full details can be found in \cite{TV06}.

Figure \ref{fig:velocity_fric}
shows a log-log plot of the  mean velocity  of an asymmetric  piston ($\alpha_+=1$,
$\alpha_-=0$) as a function of the dimensionless
dry friction $F^*$ for different mass ratios $M/m=1,2,10,20$. We observe two scaling regimes: at low dimensionless friction force $F^*$
the dimensionless mean velocity depends weakly on $F^*$, whereas in the high-friction limit $F^*>1$, $<V^*>$ decays as $F^{*-1}$.

\begin{figure}[t]
\centering
\resizebox{8cm}{!}{\includegraphics{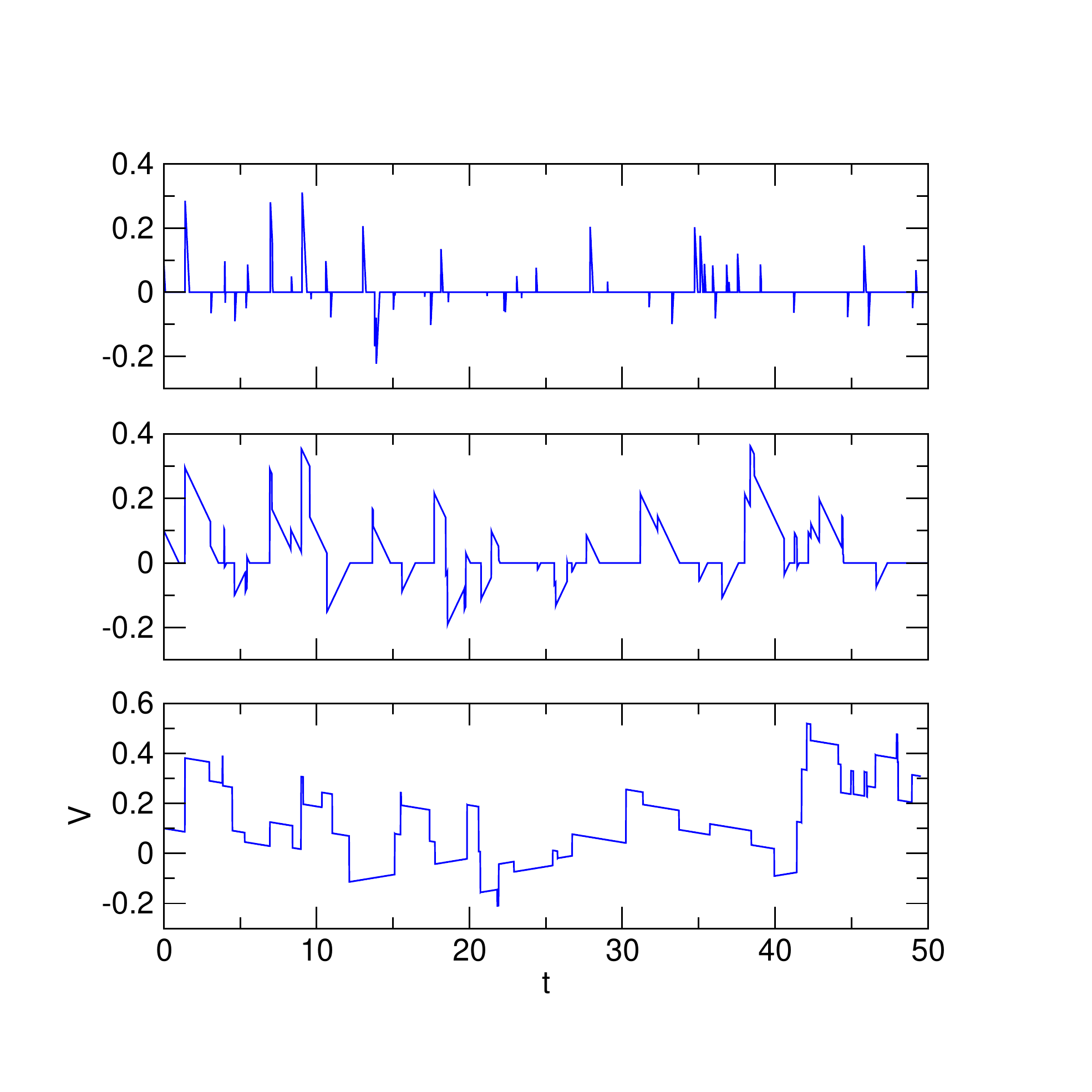}}
\caption{Dimensionless velocity, $V^*$, as a function of the reduced time $t^*=\rho L\nu(0)t$ for $F^*=1.0,0.1,0.01$ top to bottom. $\mu=10$.}
\label{fig:dynamics}
\end{figure}
\begin{figure}[t]
\resizebox{9.0cm}{!}{\includegraphics{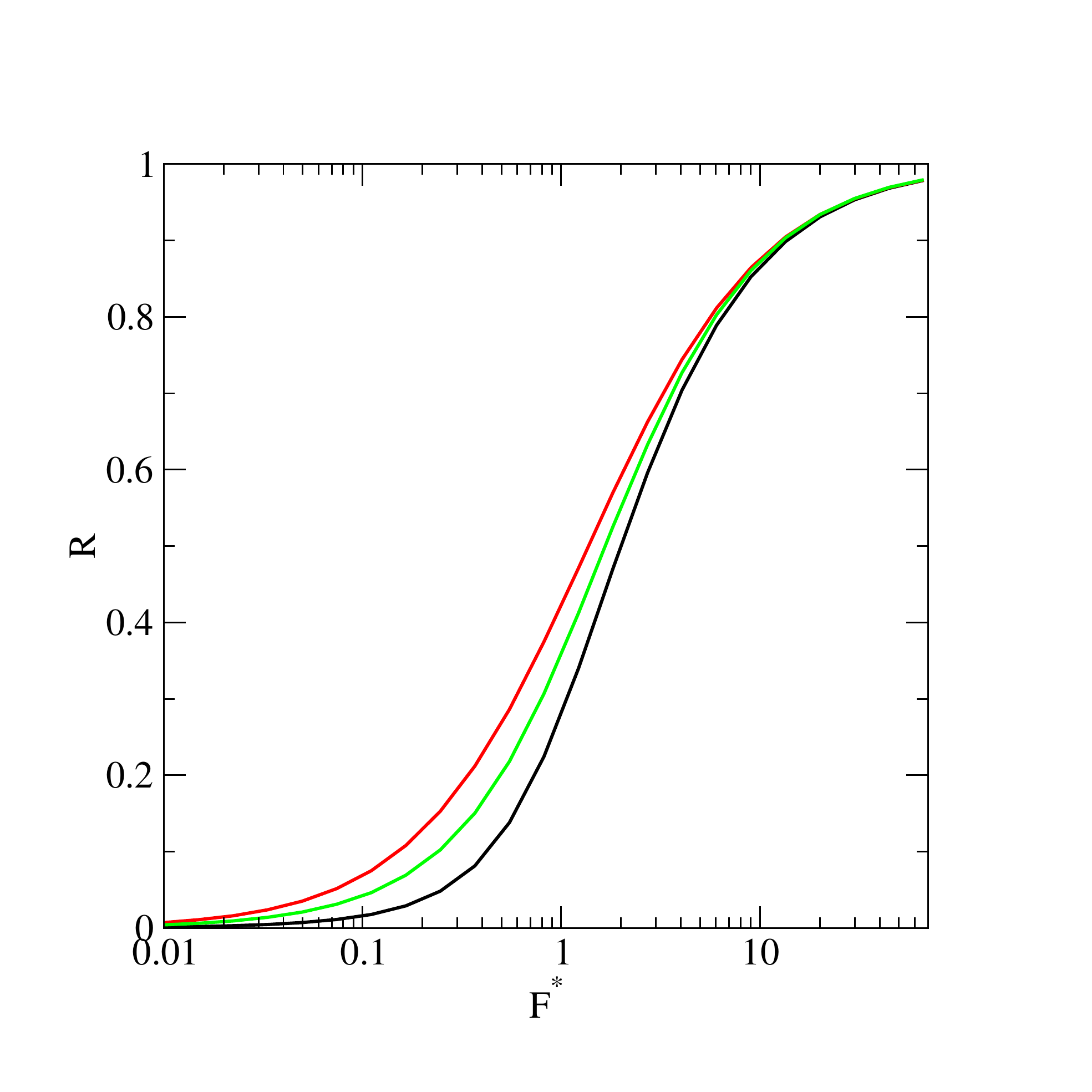}}
 \caption{(Color online) Fraction  of collisions occurring when the piston is at
rest over the number of collisions for 
different masses. From left to right: $M=2,5,20$.}
 \label{fig:ratio}
 \end{figure}

Useful insight can be obtained by observing the dynamics for different values of $F^*$: See Figure \ref{fig:dynamics}.
For $F^*=1.0$ the motor decelerates rapidly after each collision until it comes to rest. It remains motionless until it is struck by another bath particle. 
For the smallest value, $F^*=0.01$, the deceleration is weak and the 
piston is always in motion. The case $F^*=0.1$ is an intermediate case. 

We have also monitored the ratio  $R$, of   the number of collisions
occurring  when the piston is at rest to the total number collisions. 
Figure \ref{fig:ratio} shows $R$ as a function of $\Gamma_s$ for
$M/m=1,2,10,20$. As $R$ approaches one the dynamics consists of a series of independent displacements, each followed by a period of rest before the next collision with a bath particle.

\section{Time scale analysis}\label{sec:tsanalysis}
As we will show, the behavior of the system is governed by the relative values of the mean collision time and the mean stopping time. 
The mean inter-collision time between bath particles and the piston
\begin{equation}
 \tau_{c}\simeq\frac{1}{\rho L\nu(0)}
\end{equation}
is $\tau_c\simeq \sqrt{\frac{\pi m}{2T}} \frac{1}{\rho L}$ for a Gaussian bath distribution,
and the mean stopping time with friction present,
$\tau_{s}=\frac{M\overline{V}}{F}$
where $\overline{V}$ is the average  velocity after a collision.
When the dimensionless friction $F^*$ is small, $\overline{V^*}\sim (\alpha_{+}-\alpha_{-})$, 
while for $F^*>>1$,
$\overline{V^*}\sim \frac{m}{M}(\alpha_{+}-\alpha_{-})$, which gives
\begin{equation}
 \frac{\tau_s}{\tau_c}\sim \frac{\alpha_{+}-\alpha_{-}}{F^*}\left\{\begin{array}{cc}
                                    1 ,&F^{*}>>1\\
                                    \frac{M}{m}, &F^{*}<<1
                                    \end{array}
\right.
\end{equation}
This behavior is illustrated in Fig.~\ref{fig:dynamics}, where one observes that $\tau_s<<\tau_c$  
for $F^*=10$, whereas  $\tau_s>>\tau_c$    for $F^*=0.1$

Whenever the dynamics consists of successive slip-stick motions, the  velocity distribution function 
of the piston contains a regular part and a delta singularity at $V^*=0$  (where $V^*=v\sqrt{m/T}$ is the dimensionless velocity)
corresponding to the situation where the  piston is at rest for a finite time before the
next collision with a bath particle:
\begin{equation}
 f(V^*)=\gamma f_R(V^*)+(1-\gamma)\delta(V^*)
\end{equation}
where $\int d V^* f_R^*(V^*)=1$ and $\gamma$ is a constant that can be determined from conservation 
of the probability current
at $V^*=0$\cite{Touchette2010}:
\begin{equation}
(1-\gamma)\int_{-\infty}^{\infty} dv^* |v^*| \phi(v^*)=2\gamma
f_R^*(0)\frac{F^*}{\mu}
\end{equation}
giving 
\begin{equation}\label{eq:gammaexact}
\gamma^{-1}=1+2C \frac{F^* }{\mu} 
\end{equation}
with $C=2f_R^*(0)/\int_{-\infty}^{\infty} dy |y| \phi(y)$ is a numerical constant.

When $\tau_c>>\tau_s$ the  frictional force
stops the piston before the next collision,
and the motor essentially evolves by following a sequence of stick-slip
motions. Most of the time, the piston is at rest and the singular
contribution is dominant, $\gamma\simeq 1/F^*$. This regime can be
described by the independent kick model introduced below.

Conversely, when $\tau_c<<\tau_s$, collisions are so
frequent that sliding dominates the piston dynamics. For all practical purposes, the piston
never stops or stops for infinitesimal durations and $(1-\gamma)\simeq F^*$.
In this case, the dynamics is well described by
a Fokker-Planck equation for $M/m>>1$.

\section{Independent Kick model}\label{sec:ikm}
When the friction force is large, the stopping time $\tau_s$ is much shorter
than the mean time between collisions $\tau_c$. The piston dynamics is then a
sequence of uncorrelated kicks immediately followed by 	a decelerated motion
that is stopped before a next collision with a particle bath. The mean velocity
is the average over all collisions.
\begin{equation}\label{eq:vbar}
\langle V \rangle =\rho L\int_{-\infty}^{\infty} dv |v| \phi(v)
 \int_0^\tau V(t)dt
\end{equation}
where $V(t)=V_0-\frac{F\sigma(V_0)}{M}t$, $\tau=\frac{|V_0|
M}{F}$ and $V_0$ is the  velocity after a collision, which
is given by
\begin{equation}
 V_0=\frac{(1+\alpha_{+})v}{1+\mu} \mbox{\rm\hspace{0.3cm} for } v>0
\end{equation} 
and 
\begin{equation}
 V_0=\frac{(1+\alpha_{-})v}{1+\mu} \mbox{\rm \hspace{0.3cm}for } v<0
\end{equation} 
Integrating over time, one obtains
\begin{equation}
\langle V \rangle =\frac{M \rho
L((1+\alpha_+)^2-(1+\alpha_-)^2)}{2 F( 1+\mu)^2}\int_0^\infty dv v^3 \phi(v)
\end{equation}
where we have assumed that $\phi(v)$ is symmetric. With this assumption the sign of
the motor effect is independent of the form of the bath velocity distribution. 
For a Gaussian bath distribution, the dimensionless mean velocity $<V^*>$ is
given explicitly by
\begin{equation}\label{eq:vel_asym}
\langle V^* \rangle =\frac{((1+\alpha_+)^2-(1+\alpha_-)^2)\mu
}{2F^*( 1+\mu)^2}\sqrt{\frac{2}{
\pi}}
\end{equation}

\begin{figure}[t]
\resizebox{9.0cm}{!}{\includegraphics{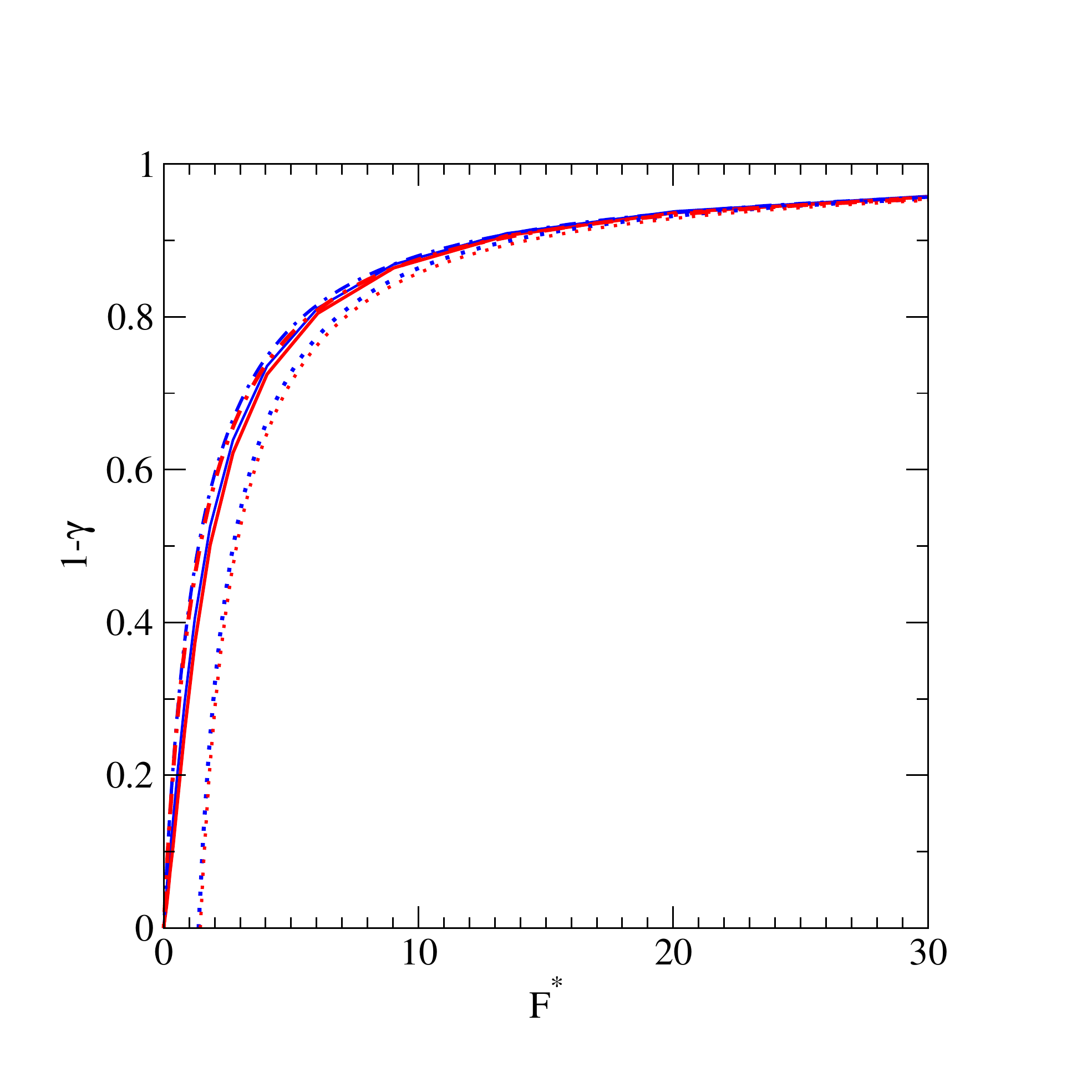}}
 \caption{(Color online) $\delta(V^*)$-contribution of  $f(V^*)$, $1-\gamma$, 
of an asymmetric  granular piston versus $F^*$:  $\alpha_+=1$, $\alpha_-=0$, for 
$M/m=10,20$.  The dotted curves correspond to the analytical expression of the
kick model and the dot-dashed curves to Eq. \ref{eq:gammaexact} where $C$ is calculated from  the exact expression in the high friction limit.}
 \label{fig:integral}
 \end{figure}

A second quantity of interest is the integral of the velocity
distribution (obtained from Eq. (\ref{eq:vbar}) by setting $V=1$).
\begin{align}\label{eq:integralfv}
I_f &=\rho L
\int_0^\infty dv v^2 \frac{|V_0|M}{F}
\phi(v)\nonumber\\
&=\frac{\rho L(2+\alpha_+ +\alpha_-)
M}{F(1+\mu)}\int_0^\infty dv v^2 \phi(v)
\end{align}
For a Gaussian bath distribution, one has
\begin{equation}\label{eq:intreg}
I_f = \frac{ (2+\alpha_++\alpha_-)\mu}{2
F^*(1+\mu)}
\end{equation}
This quantity corresponds to the value of $\gamma$ in the limit of large friction (or small ratio
$\tau_s/\tau_c$). Figure \ref{fig:integral} shows the $\delta(V^*)$-contribution of  $f(V^*)$, $1-\gamma$ for
two mass ratios $M/m=10,20$. The exact expression of the kick model Eq. (\ref{eq:intreg}) (dotted curves) underestimates the  
$\delta(V^*)$-contribution of  $f(V^*)$. The dot-dashed curves correspond to the exact expression, Eq. (\ref{eq:gammaexact}),
where $C$ is calculated by performing an exact asymptotic expansion of Eq. (\ref{eq:integralfv}) and  matching  with the independent kick model 
 Eq. (\ref{eq:gammaexact}). In this case a better agreement with simulations for small $F^*$ is observed.  
The remaining small excess is due to the fact $C$ is set to a 
constant, but in reality depends slightly on $F^*$.

The characteristic function can be also calculated
\begin{equation}
\langle e^{i k V} \rangle =\rho L\int_{-\infty}^{\infty} dv |v| \phi(v)
 \int_0^\tau e^{i k V(t)}dt
\end{equation}
Integrating over time, one obtains
\begin{align}
\langle e^{i k V } \rangle &=\frac{\rho L M}{i k F} \int_{0}^{\infty} dv
v \phi(v)
  \nonumber\\
&\left[\exp\left(\frac{ik(1+\alpha_+)
 v}{1+\mu}\right)-\exp\left(\frac{-ik(1+\alpha_-) v}{1+\mu}\right)\right]
\end{align}
Taking the inverse Fourier transform,  one infers the velocity distribution
$f_R(V)$
\begin{align}
\gamma f_R(V)&=\frac{\mu\rho T}{F}\left[\theta(V)\int_{\frac{(1+\mu) V}{1+\alpha_+}}^\infty dv |v|\phi\left(v
\right)\right.\nonumber\\ 
&\left.+\theta(-V)\int^\frac{(1+\mu)V}{1+\alpha_-}_{-\infty} dv |v| \phi\left(v\right) \right]
\end{align}
Note that the regular  velocity distribution $f_R(V)$ is continuous at $V=0$.
For a Gaussian  bath distribution, the dimensionless velocity distribution is then given by

\begin{align}\label{eq:distribution}
 \gamma f_R(V^*)&=\frac{\mu}{F^*}\sqrt{\frac{1}{2\pi}}\left[\theta(V)\exp\left(-\frac {(1+
\mu)^2 V^{*2}}{2 (1+\alpha_+)^2  }\right)\right.
\nonumber\\&\left.+\theta(-V^*)\exp\left(-\frac {(1+\mu)^2
  V^{*2}}{2(1+\alpha_-)^2  }\right)\right]
\end{align}

\begin{figure}[t]
\resizebox{9.0cm}{!}{\includegraphics{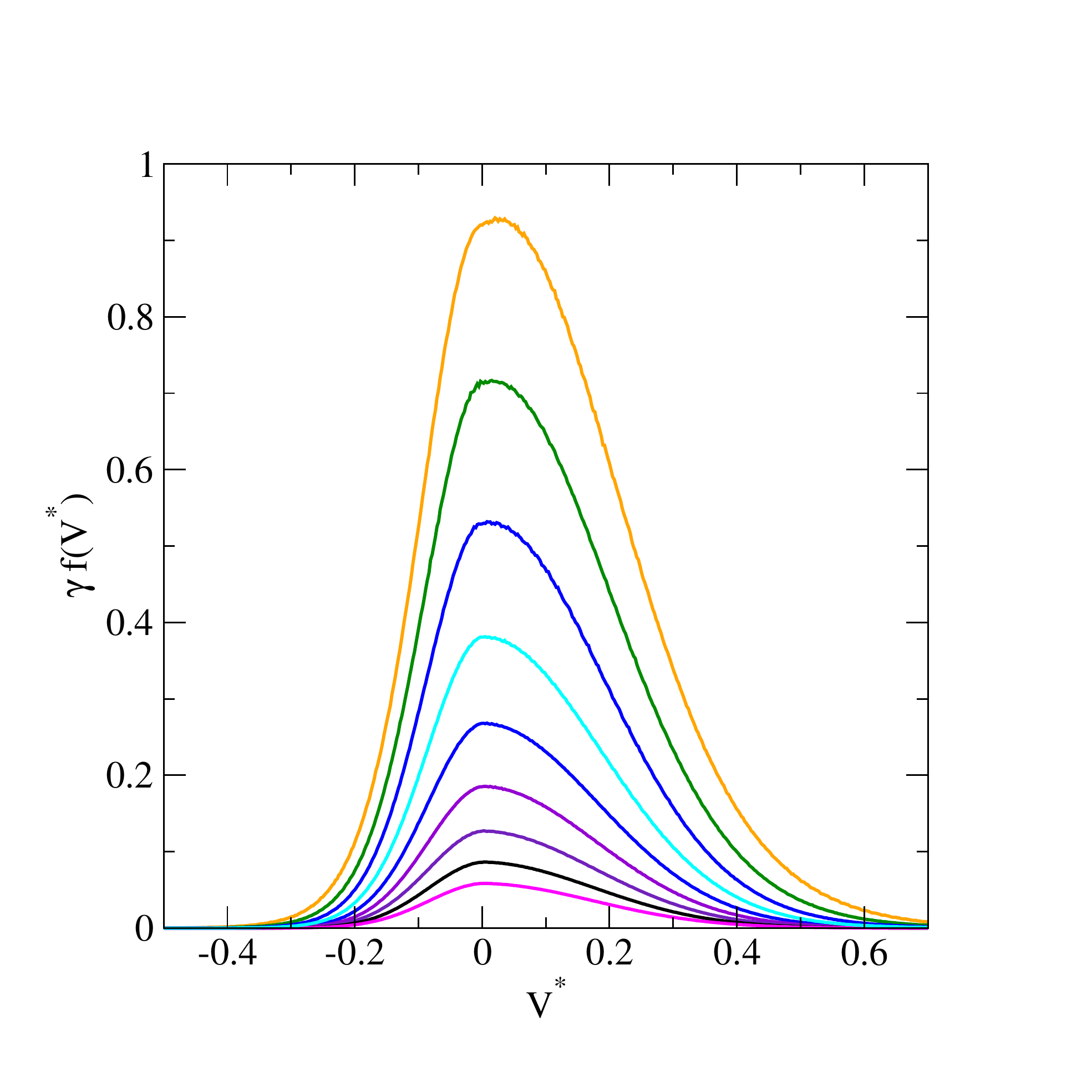}}
 \caption{(Color online) Dimensionless Velocity distributions $ \gamma   f(V^*)$ 
for various dimensionless friction forces
$F^*=2.72,6,05,6.6,9.02,13,6,20.09,30.0,44.7,66.7$ (From top to bottom). }
 \label{fig:distribM10}
 \end{figure}

\begin{figure}[t]
\resizebox{9.0cm}{!}{\includegraphics{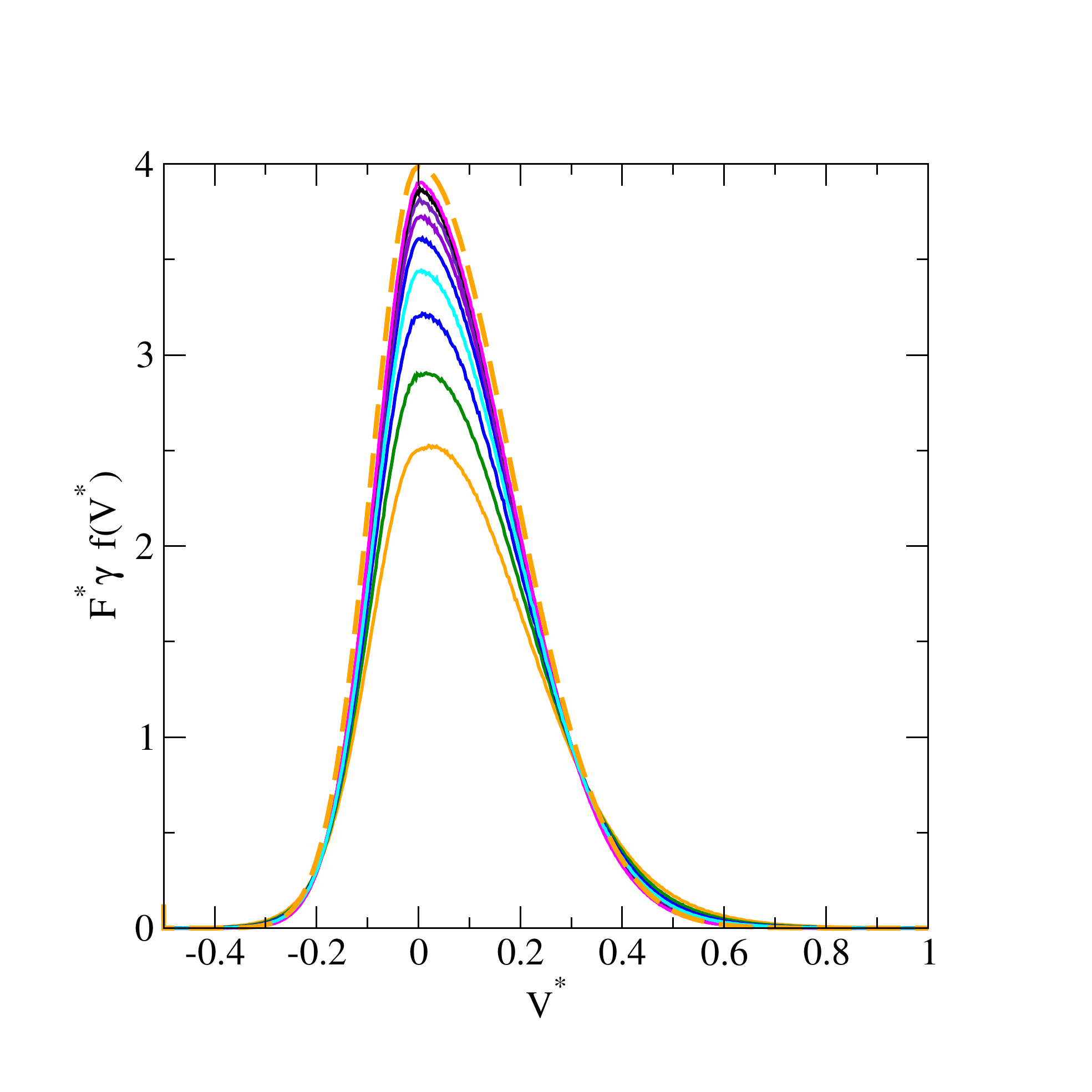}}
 \caption{(Color online) Rescaled velocity distributions $ F^* \gamma f(V^*)$ for the same
values of $F^*$ as in Fig. \ref{fig:distribM10} but in reverse order.}
 \label{fig:distribM10_scale}
 \end{figure}

Velocity distributions are displayed in
Fig.\ref{fig:distribM10} for different values of the solid friction. 
As expected the amplitude decreases as this quantity increases. 
Rescaled distributions  $F^* f_R(V^*)$ versus $V^*$ are 
shown in Fig. \ref{fig:distribM10_scale} where one observes  
 that for $F^*>2$, curves converge towards the exact result,
Eq. (\ref{eq:distribution}).

\section{The Brownian limit and the Fokker-Planck equation}\label{sec:BLFPE}
We now consider the opposite limit when the stopping time $\tau_s$ is much larger than 
the mean time between bath collisions $\tau_c$.
 For homogeneous granular motors where the mean  velocity goes to zero in the Brownian limit
 \cite{Brey1999,Cleuren2007},  a standard Kramers-Moyal (KM) expansion of the BL
integral operator leads to the Fokker-Planck differential operator. 
For  heterogeneous granular motors, where the mean velocity remains finite and independent of the mass ratio in the Brownian limit,
some caution is needed: The perturbative expansion must be performed around the non-zero mean velocity (rather than around zero)\cite{Talbot2011}. 
Finite mass corrections, however, cannot be easily included.

The approach we proposed in \cite{PhysRevLett.107.138001} 
consists of reexpressing the BL operator as a complete series expansion in terms of the derivatives of the velocity distribution function. Applying the same method to the piston we have, e.g. 
\begin{equation}
 f\left(V-\frac{1+\alpha_{+}}{1+\mu}y\right)=\sum_{n=0}^{\infty}\left(\frac{1+\alpha_+}{1+\mu}\right)^n \frac{(-y)^n}{n!}\frac{\partial^n f(V)}{\partial V^n}
\end{equation}
with similar expressions for  $\phi(V+y-\frac{1+\alpha_{+}}{1+\mu}y)$, $f(V+\frac{1+\alpha_{-}}{1+\mu}y)$ and $\phi(V-y+\frac{1+\alpha_{-}}{1+\mu}y)$. Inserting in the BL equation
 (\ref{eq:BLy})  allows us to write the collision operator as
\begin{equation}\label{eq:BLo}
 J[f,\phi]= \sum_{n=1}^\infty
\frac{1}{n!\mu^n}\frac{\partial^n(g_n(V)f(V))}{\partial V^n}
\end{equation}
where we have used the fact that the zero order terms of the expansion cancel the destruction term and where we have introduced
\begin{align}
g_n(V)&=\rho L\int_0^{\infty}dy\;y\left[\left(-\mu\frac{1+\alpha_+}{1+\mu}\right)^n\phi(V+y)+\right.\nonumber\\
&\left.\left(\mu\frac{1+\alpha_-}{1+\mu}\right)^n\phi(V-y)\right]
\end{align}
The functions $g_n(V)$ can be obtained from the generating function
\begin{align}
 &g(V,a)=\rho L \int_0^\infty dy y
\left[\exp\left(\frac{-(1+\alpha_{+})\mu ya}{1+\mu}\right)\right.
 \phi(V +y) \nonumber\\
& +\left. \exp\left(\frac{(1+\alpha_{-})\mu ya}{1+\mu}\right)\phi(V
-y)\right]
\end{align}


Truncating the BL operator, Eq. (\ref{eq:BLo}), at  second-order and
adding the dry friction leads to the following Fokker-Planck equation
\begin{align}\label{eq:FP}
\frac{\partial f(V,t)}{\partial t}&=\frac{1}{M}
\frac{\partial}{\partial V}\left[(F\sigma(V)+mg_1(V)) f(V,t)\right]
\nonumber\\&+\frac{1}{2M^2}\frac{\partial^2}{\partial
V^2}[m^2g_2(V)f(V,t)]=0
\end{align}
in which all finite-mass   
corrections are incorporated,  and where deviations from a Gaussian distribution are present for
large finite masses. Recalling  that the $g_n(V)$ are proportional to $\rho$, we see that for a given dry friction $F$,
increasing the bath density reduces the effect of friction.
The corresponding Langevin equation \cite{Gardiner_2009} features a motor force with a
non-linear dependence on $V$ and a colored noise.

\begin{equation}
 M\frac{d V}{dt}=-F\sigma(V)-mg_1(V)+m\sqrt{g_2(V)}\eta(t)
\end{equation}
where $\eta(t)$ is a white Gaussian noise, where $\langle\eta(t)\rangle=0$ and $\langle \eta(t)\eta(t')\rangle=\delta(t-t')$

The steady state solution of Eq. (\ref{eq:FP}) is
\begin{equation}
 f(V)=\frac{C_{f}}{g_2(V)}\exp\left[-2M\int_0^V du\; 
\left(\frac{mg_1(u)+F\sigma(u)}{ m^2g_2(u)}\right)\right]
 \end{equation}
where $C_{f}$ is obtained from the normalization condition $\int d V
f(V)=1$. This result clearly shows that, even in the absence of friction,
the velocity distribution is non-Gaussian for finite mass ratios.

In the Brownian limit $g_1(V)\sim g'_1(\tilde{V})(V-\tilde{V})$ and
$g_2(V)=2T_g/m g'_1(\tilde{V})$, where $T_g$ (the  granular temperature of the piston 
which is lower than the bath temperature, $T$) and $\tilde{V}$ are given by the Kramers-Moyal
expansion (Eqs. (6) and (8) in Ref.~\cite{PhysRevE.82.011135}). (Physically, $\tilde{V}$ is the exact steady state drift velocity of a piston in the Brownian limit in the absence of friction). 
This finally gives a stationary
distribution, at the lowest order in $m/M$,
\begin{equation}\label{eq:FPF}
 f(V)=C \exp\left(-\frac{M(V-\tilde{V})^2}{2T_g}-\frac{ \mu F |V|
} {g_1'(\tilde{V})T_g}\right)
\end{equation}
where $C$ is the normalization constant.
Whereas  one observes a Gaussian  decay of the velocity distribution at large velocity,
$f(V)$ decreases  exponentially for small and intermediate velocities, due to
friction \cite{JPSJ.73.2037,Chaudhury2008,Touchette2010}.

%

\section{Static Friction}\label{sec:static}

Our analysis has so far been restricted to dynamic dry friction that produces a constant
retarding force on a moving piston. A stationary piston acquires a non-zero
velocity following a collision with a bath particle, no matter how slowly the latter is moving. 
In reality, static friction will also be present and this will prevent the piston from moving unless it is struck by a sufficiently fast moving
bath particle. To model this effect correctly using a coefficient of static friction we would need
to know the time-dependent force acting on a stationary piston during a collision with a bath particle. If this
force exceeds the force due to static friction, the piston starts to move. 
In the present model, however, the collisions are assumed to be instantaneous. 
Therefore, we represent static friction in an approximate way, by introducing 
a threshold impulse $I_m$ so that the piston is only set into motion if  $|I_{\pm}|>I_m$.


\begin{equation}\label{eq:static}
 V' = \left\{ \begin{array}{rl}\displaystyle
 \frac{m}{m+M}(1+\alpha_{+})v-\frac{I_m}{m+M} &\mbox{ if $v>\frac{I_m}{m(1+\alpha_{+})}$} \\
\displaystyle \frac{m}{m+M}(1+\alpha_{-})v+\frac{I_m}{m+M} &\mbox{ if $v<-\frac{I_m}{m(1+\alpha_{-})}$} \\\displaystyle
  0 &\mbox{ otherwise}
       \end{array} \right.
\end{equation}
This implies that a colliding bath particle must be moving faster than
\begin{equation}
 v^{\pm}_m=\pm\frac{I_m}{m(1+\alpha_\pm)}
\end{equation}
to set a stationary piston into motion. 
Let us introduce the dimensionless threshold impulse
\begin{equation}
 I^*_m=\frac{I_m}{\sqrt{mT}}
\end{equation}
the effect of static friction consists of adding a singular contribution to the
velocity distribution and of weakening the motor effect.

When the time scale ratio $\tau_s/\tau_c>>1$, it is easy to generalize the
independent kick model by adding the static friction.
The drift velocity is given by 
\begin{align}
 &\langle V \rangle =\rho L\int_{v^+_m}^\infty dv v \phi(v)
\int_0^\tau dt V(t) \nonumber\\
&+\rho L\int^{v^-_m}_{-\infty} dv v \phi(v)
\int_0^\tau dt V(t)
\end{align}
where $V(t)=V_0-\frac{F\sigma(V)}{M}t$, $\tau=\frac{M|V_0|}{F}$ and $V_0$ is
given by Eq.~(\ref{eq:static}).

Integrating over time, one obtains the following expression
\begin{align}
 &\langle V \rangle =\frac{M\rho
L}{2F(1+\mu)^2}\left((1+\alpha_+)^2\int_{v^+_m}^\infty dv \,v (v-v^{+}_m)^2 \phi(v)
 \right.\nonumber\\
&\left.-(1+\alpha_-)^2\int_{-\infty}^{v^-_m}dv \,|v| (v^{-}_m-v)^2 \phi(v)\right)
\end{align}
For a Gaussian bath velocity distribution, the dimensionless mean velocity is given by
\begin{align}\label{eq:staticV}
 \langle
&V^*\rangle=\frac{\mu}{2F^*(1+\mu)^2}
[h(I^*_m,\alpha_+)-h(I^*_m,\alpha_-)]
\end{align}
with
\begin{align}
h(I^*_m,\alpha)=(1+\alpha)^2\sqrt{\frac{2}{\pi}}\exp\left(-\frac{I_m^{*2}}{2(1+\alpha)^2}\right) \nonumber\\
 -(1+\alpha)I_m^{*}erfc \left( \frac{I_m^{*}}{\sqrt{2}(1+\alpha)}\right)
\end{align}
that correctly reduces to Eq. (\ref{eq:vel_asym}) if $I^*_m=0$. 
\begin{figure}[t]
\resizebox{9.0cm}{!}{\includegraphics{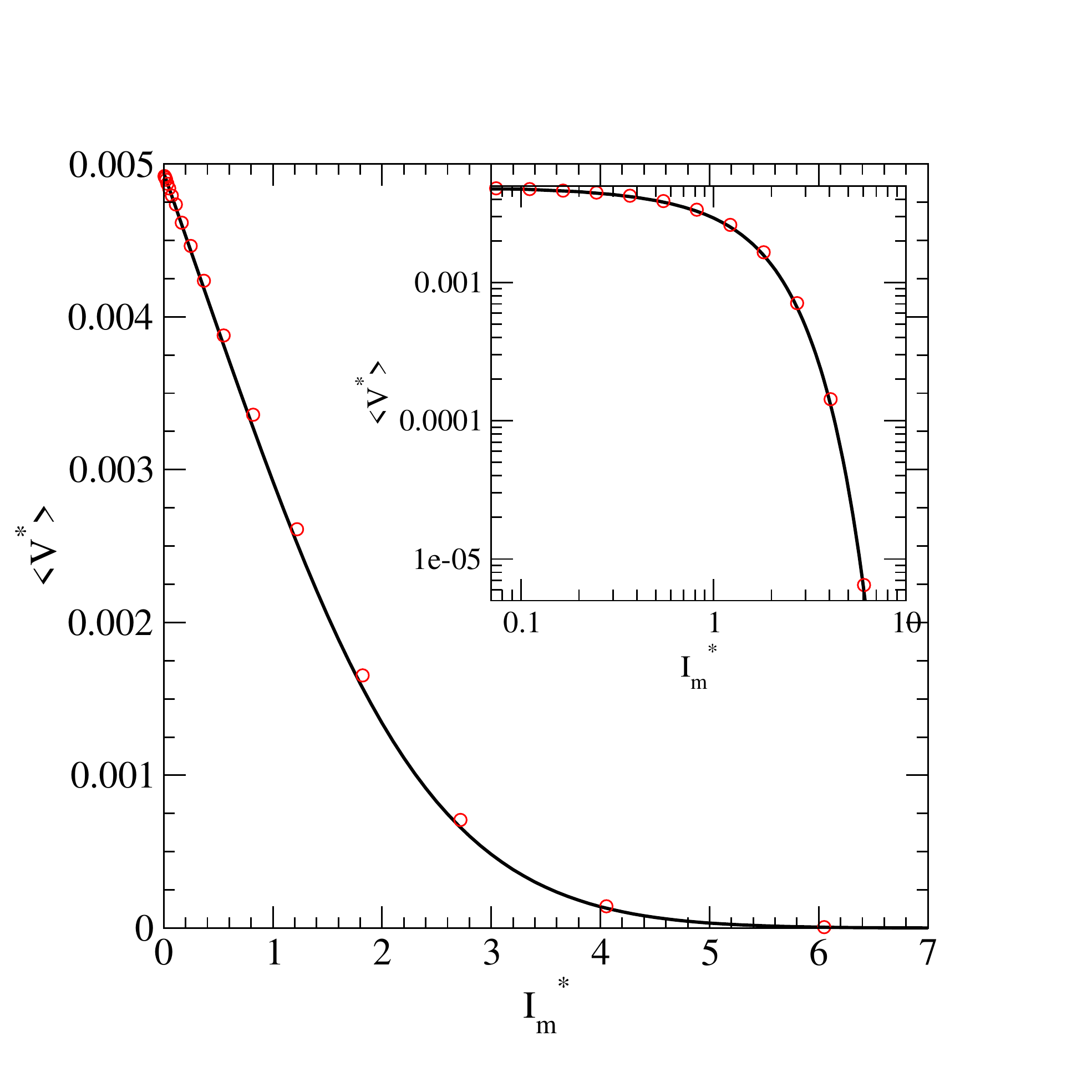}}
 \caption{(Color online) Dimensionless  mean velocity    of 
an asymmetric  granular piston of mass ratio $M/m=10$ with a friction force $F^*=20.01$: 
$\alpha_+=1$, $\alpha_-=0$, versus the threshold impulse.   The circles  correspond to the simulation results  with the static friction and the solid lines show the theoretical result, Eq. (\ref{eq:staticV}).}
 \label{fig:staticV}
 \end{figure}
Fig. \ref{fig:staticV} shows that the theoretical result is in good agreement with numerical simulations of the Boltzmann-Lorentz equation with static and dynamic friction. 

Developing Eq. (\ref{eq:staticV}) as a power series about $I^*_m=0$, we obtain
\begin{align}
 \langle V^*\rangle=\frac{\mu}{2F^*(1+\mu)^2}(\alpha_+-\alpha_-)\left[(2+\alpha_++\alpha_-)\sqrt{\frac{2}{\pi}}\right.\nonumber\\
\left.-I^*_m+O(I^{*4}_m)\right]
\end{align}
We note the absence of terms in $I^{*2}_m$ and $I^{*3}_m$ and from Fig. \ref{fig:staticV} we see that the mean velocity increases linearly for $I^*_m<2$. 
Furthermore, a non-zero impulse threshold means that, in additon to the simple scaling behaviour ($\langle V\rangle\sim T^{3/2}$) with only dynamic friction present, a sub-dominant term appears:
\begin{equation}
 \langle V\rangle \propto (\alpha_{+}-\alpha_{-})\left[(2+\alpha_{+}+\alpha_{-})\sqrt{\frac{2}{\pi}}T^{3/2}-\frac{I_m}{\sqrt{m}}T\right]
\end{equation}

The asymptotic behavior for large values of $I^*_m$ is
\begin{align}
 \langle V^*\rangle&=\frac{\mu}{2F^*(1+\mu)^2}\frac{1}{I^{*2}_m}\sqrt{\frac{2}{\pi}}\left[(1+\alpha_+)^4\exp(-\frac{I^{*2}_m}{2(1+\alpha_+)^2})\right.\nonumber\\
&\left.-(1+\alpha_-)^4\exp(-\frac{I^{*2}_m}{2(1+\alpha_-)^2})\right]
\end{align}
In this limit only fast-moving bath particles, which are few since they correspond to the tails of the bath velocity distribution, can actuate a stationary piston. Therefore, the motor effect vanishes. 
The integral of the velocity distribution is
\begin{align}
I_f=\frac{ML\rho}{(1+\mu) F}\left[(1+\alpha_+)\int_{v^+_m}^\infty dv v (v-v^+_m)
\phi(v)\right. \nonumber\\\left.
+(1+\alpha_-)\int^{v^-_m}_{-\infty} dv |v| (v^-_m-v) \phi(v)\right]
\end{align}
that, for a Gaussian bath velocity distribution, gives

\begin{align}
I_f=&\frac{\mu}{(1+\mu)F^*}\left[(1+\alpha_+) erfc \left(\frac{I^*_m}{\sqrt{2}(1+\alpha_+)}
\right)\right. \nonumber\\&\left.
+(1+\alpha_-)erfc \left(\frac{I^*_m}{\sqrt{2}(1+\alpha_-)}
\right)\right]
\end{align}
that goes rapidly to zero for
$I^*_m>2$. 
\begin{figure}[t]
\resizebox{9.0cm}{!}{\includegraphics{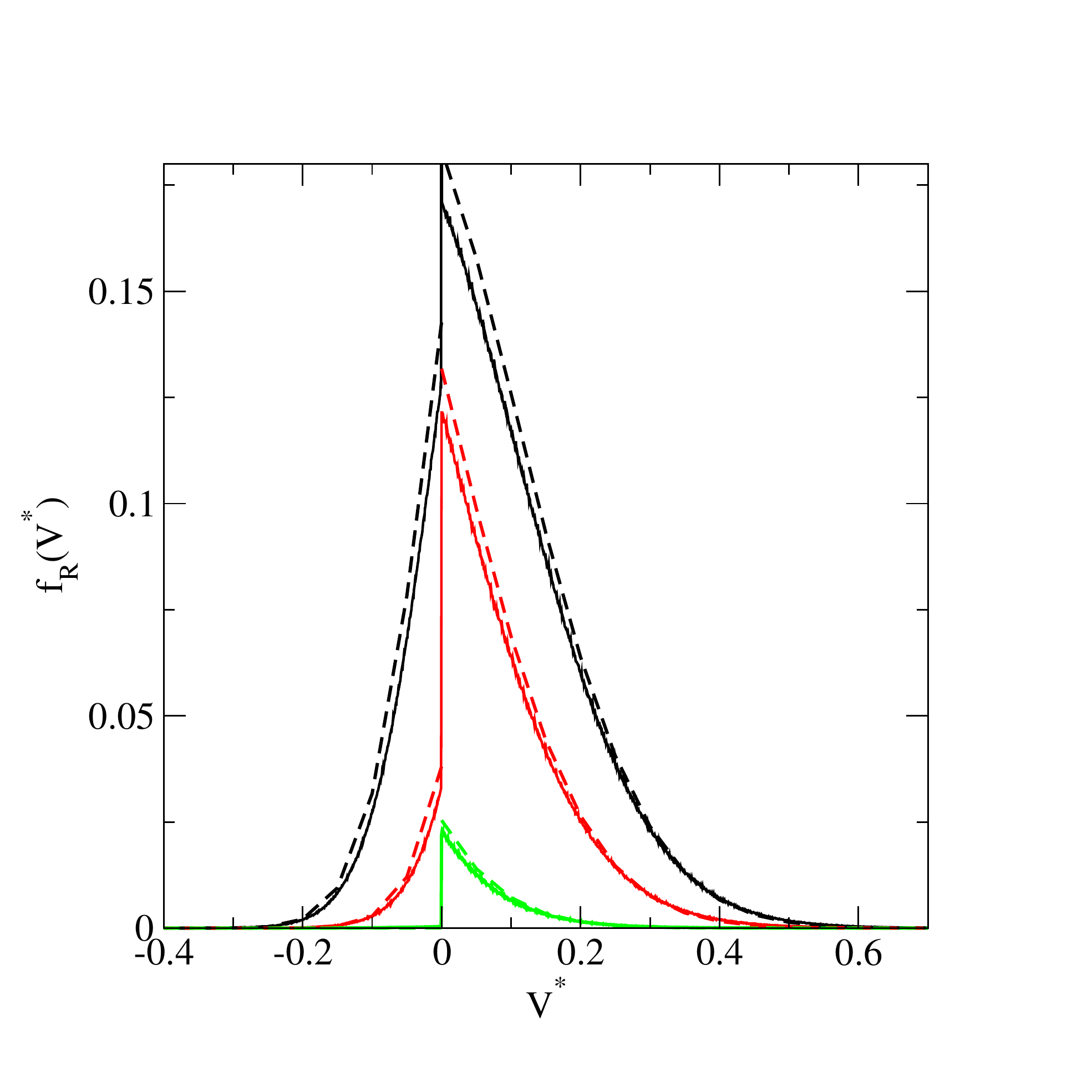}}
 \caption{(Color online) Dimensionless  velocity distributions   of 
an asymmetric  granular piston of mass ratio $M/m=10$ a friction force $F^*=20.01$: 
$\alpha_+=1$, $\alpha_-=0$, with different static frictions 
$I_m^*=0.82,1.82,4.06 $.  The dashed curves correspond to the analytical expression
of the independent kick model with the static and dynamic frictions,  Eq. (\ref{eq:staticVD}).}
 \label{fig:static10}
 \end{figure}

Finally, the characteristic function $\langle e^{ikV}\rangle$ is given by
\begin{align}
 &\langle e^{ikV}\rangle=
\frac{\rho ML}{F}
\left[\int_{v^+_m}^\infty dv |v| \phi(v) \frac{e^{i\frac{1+\alpha_+}{1+\mu}k(v-v^+_m)}-1}{ik}\right.\nonumber\\
&\left.+\int^{v^-_m}_{-\infty} dv |v| \phi(v) \frac{e^{i\frac{1+\alpha_-}{1+\mu}k(v-v^-_m)}-1}{ik}\right]
\end{align}

Taking the inverse Fourier transform one obtains the velocity distribution
\begin{align}
 &\gamma f_R(V)=
\frac{\rho ML}{F}
\left[\theta(V)\int_{\frac{I^*_m}{1+\alpha_+}+\frac{1+\mu}{1+\alpha_+}V}^\infty dv |v| \phi(v)\right.
\nonumber\\
&+\theta(-V)\left.\int^{\frac{-I^*_m}{1+\alpha_-}+\frac{1+\mu}{1+\alpha_-}V}_{-\infty} dv |v| \phi(v) \right]
\end{align}

For a Gaussian distribution, the dimensionless velocity distribution is expressed as
\begin{align}\label{eq:staticVD}
 &\gamma f_R(V^*)=
\frac{ \mu}{F^*}\sqrt{\frac{1}{2 \pi}}\left[\theta(V) \exp\left(-\frac{\left({I_m^*+(1+\mu)V^*}\right)^2}{2(1+\alpha_+)^2}\right)
\right.
\nonumber\\
&+\left. \theta(-V)\exp\left(-\frac{\left({-I_m^*+(1+\mu)V^*}\right)^2}{2(1+\alpha_-)^2}\right)  \right]
\end{align}
We see that static friction has a dramatic qualitative effect on the regular velocity distribution in the sense that it is  discontinuous at $V^*=0$, 
a feature  not present for $I^*_m=0$.


\begin{align}
 \gamma f_R(0+)-\gamma f_R(0-)=\frac{ \mu}{F^*}\sqrt{\frac{1}{2 \pi}}
\left[ \exp\left(-\frac{I_m^{*2}}{2(1+\alpha_+)^2}\right)
\right.\nonumber\\ \left. -
\exp \left(-\frac{I_m^{*2}}{2(1+\alpha_-)^2}\right)
\right]
\end{align}

Figure \ref{fig:static10} shows velocity distributions when $F^*=20.01$, $M/m=10$, and 
for different values of the impulse threshold (full curves). The independent kick model incorporating dynamic and static friction 
provides an accurate description of the kinetic properties of the Boltzmann-Lorentz equation. As expected, one observes a finite discontinuity of
the velocity distribution at $V=0$. Negative piston velocities occur less frequently  than positive ones due to the low efficiency
of bath collisions on the right side of the piston ($\alpha_-=0$) and  the momentum threshold of the static friction. As the static friction increases,
motion of the piston to the left is reduced more rapidly than motion to the right.

\section{Conclusion}\label{sec:conclusion}
We have considered the effect of dynamic and static friction on the kinetics of a granular asymmetric piston.
When only dynamic friction is present, the mean velocity exhibits two scaling regimes  depending on the strength of the friction force.
In the high friction limit, the Boltzmann-Lorentz equation is asymptotically described by a solvable independent kick model. Conversely, 
when the friction is small, and if the mass ratio is large, the model can be mapped to Fokker-Planck equation, for which exact results 
can be also  obtained. When static, as well as dynamic, friction is present the mean piston velocity initially decreases linearly with the threshold impulse, 
while for larger values of this parameter the motor effect is rapidly suppressed, decreasing in a Gaussian fashion.

Further investigation could consider collective effects of motor assemblies observed in biological systems
\cite{Howard2009,PhysRevLett.104.248102,PhysRevLett.106.068101} or dense granular systems with active particles 
\cite{PhysRevE.81.061916,PhysRevLett.101.268101,PhysRevLett.105.098001,Baskaran2010}. 

\bibliographystyle{apsrev4-1}       

%
\end{document}